\documentclass[pra,aps,a4paper,showpacs,superscriptaddress,twocolumn]{revtex4}
\usepackage{mathrsfs}
\usepackage{amsfonts}
\usepackage{amstext}
\usepackage{amsmath}
\usepackage{amssymb}
\usepackage[dvips]{graphicx}

\begin{document}

\title[Short title for running header]{ Diffractions of Bose-Einstein
Condensate in Quantized Light Fields}

\author{Peng Zhang}
\affiliation {Beijing National Laboratory for Condensed Matter
Physics, Institute of Physics, Chinese Academy of Sciences, Beijing
100190, China}

\author{Z.-Y. Ma}
\affiliation{Shanghai Institute of Optics and Fine Mechanics,
Chinese Academy of Sciences, Shanghai 201800, China}

\author{Jian-Hua Wu}
\affiliation {Beijing National Laboratory for Condensed Matter
Physics, Institute of Physics, Chinese Academy of Sciences, Beijing
100190, China}

\author{H. Fan}
\affiliation {Beijing National Laboratory for Condensed Matter
Physics, Institute of Physics, Chinese Academy of Sciences, Beijing
100190, China}

\author{W. M. Liu}
\affiliation {Beijing National Laboratory for Condensed Matter
Physics, Institute of Physics, Chinese Academy of Sciences, Beijing
100190, China}

\date\today

\begin{abstract}
We investigate the atomic diffractions of a Bose-Einstein condensate in quantized light fields. Situations in which the light fields are in number states or coherent states are studied theoretically. Analytical derivation and numerical calculation are carried out to simulate the
dynamics of the atomic motion. In condition that atoms are scattered by light
in the number states with imbalanced photon number distribution, the
atomic transitions between different momentum modes would
sensitively depend on the transition order and the photon number
distribution. The number-state-nature of the light fields modifies
the period of atomic momentum oscillations and makes forward and
backward atomic transitions unequal. For light fields in coherent
states, no matter the intensities of the light fields are balanced
or not, the atomic diffractions are symmetric and independent on the
transition order.
\end{abstract}

\maketitle

\address {Beijing National Laboratory for Condensed Matter Physics,
Institute of Physics, Chinese Academy of Sciences, Beijing 100190,
China}

\section{Introduction}
The coherent interaction between matter and electromagnetic fields
within various kinds of physical system is providing a useful
platform for developing concepts in quantum optics and atomic
molecular physics. The experimental realization of Bose-Einstein
condensation in dilute atomic gases \cite{BECP}, greatly facilitates
the investigations of the interactions between the atom and light
on a macroscopic scale
\cite{Anderson,Cataliotti,wavemixing,JMZhang,GDLin}. In the past few
years, atomic scattering process in optical fields has been
intensively studied, such as matter-wave supperradiance
\cite{Inouye,Zhou,Moore}, quantum phase transition
\cite{Baumann,chin}, cavity optomechanics \cite{Brennecke,Aranya},
raising a variety of striking discoveries.

In physical systems involving light-atom interactions, various kinds
of experimental configurations have been developed or proposed to
achieve the desired atom-field coupling \cite{Li Ke, zhang}.
Especially, by using high-quality resonators
\cite{SHuang,GXLi,Horak}, strong coupling regime of the light-atom
can be reached, where atoms coherently exchange photons with light
fields. However, in the previous light-atom interacting models, the
derivations usually use the classical treatment for the
electromagnetic fields, in which the light fields are recognized as
amplitude-modulated plane waves with slowly varying amplitudes; or
even in a quantized treatment for light fields, mean-field
approximations is often introduced for the light fields. Atomic
diffractions in such situations have been well studied, however, in
most conditions, the subtle effects of the atomic motion induced by
the quantum nature of light has always been ignored.

In this paper, we study the diffractions of a Bose-Einstein
condensate (BEC) in multimode optical fields, with a full quantum
treatment of the light fields. In the cases that the quantized light
fields are in different kinds of states, exotic behaviors are
induced in the atomic motion by the quantum-nature of light. This
paper is organized as following, in section II, we introduce the
physical system and develop the theoretical model of the atomic
diffractions in a full quantum treatment of the light fields. In
section III, we analyze how the quantum status of the pump fields
can modify the atomic motion in the cases that the optical fields
are in number states. In section IV, we turn to the case where the
light fields are in coherent states. In section V, we arrive at a
summary of our results.

\section{the theoretical model for atomic diffraction in quantized light fields}

\begin{figure}[!h]
\begin{center}
\scalebox{0.40}[0.38] {\includegraphics*
[160pt,430pt][590pt,790pt]{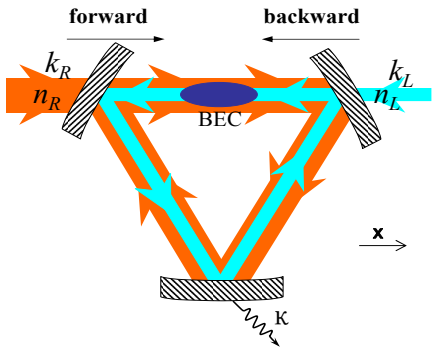}}
\end{center}
\caption{(Color online)  A schematic diagram of a two-mode triangle
ring cavity containing
 an elongated Bose-Einstein condensate trapped along one of its cavity axis. The
 damping rate $\kappa$ of the cavity modes is small compared to the atomic motion inside
 the cavity. There are two degenerate light modes propagating in the clockwise
  and anticlockwise directions inside the cavity.}
\label{setup}
\end{figure}

The physical system under study is illustrated in Fig.~\ref{setup},
where an elongated BEC is trapped along the horizontal axis of a
triangle ring cavity. The cavity consists of three mirrors which are
placed in a way such that the light reflected inside the cavity
forms a closed loop. There are two mutually counterpropagating
modes, i.e., the clockwise and anti-clockwise modes, coupled equally
to the atoms. The light fields in these two modes are also referred
as the forward (or right-going, with wave-vector$k_{R}$) and
backward (or left-going, with wave-vector $k_{L}$) ones according to
the geometrical configuration. These two cavity modes are
degenerated in frequency, $\omega _{R}=\omega _{L}=\omega _{c}$, so
the wave-vectors of the two cavity mode satisfies $k_{R}=-k_{L}$.
The atoms within the condensate are recognized as two-level atoms
with one internal ground state $ \left\vert g_{in}\right\rangle $
and excited state $\left\vert e_{in}\right\rangle $ representing the
internal degrees of freedom for each atom. In the large detuning
limit, the upper internal state $\left\vert e_{in}\right\rangle $\
of the atoms can be adiabatically eliminated
\cite{Dimer,Moore,Ozgur}. For illustration, the condensate is chosen
to be prepared with $^{87}$Rb atoms, and the wave length of the
light fields is $\lambda=780$nm with a detuning
$\Delta=\omega_{c}-\omega_{a}=-1.5$GHz of the $D_{2}$ transition
\cite{Zhou, Li Ke, zhang}.
The Hamiltonian of the light-atom coupling system can be written as,
\begin{eqnarray}
\hat{\mathscr{H}}\!\!\! &=&\!\!\!\int \textrm{d}^{3}\textrm{r}\hat{\Psi}^{\dag }(\textrm{\textbf{r}})\Big[-\frac{
\hbar ^{2}}{2M}\nabla ^{2}+\hat{\mathbf{E}}(\textrm{\textbf{r}})\cdot \hat{\mathbf{D}}(\textrm{\textbf{r}})\Big]%
\hat{\Psi}(\textrm{\textbf{r}})  \notag \\
&+&\!\!\!\frac{1}{2}\!\!\int
\!\!\textrm{d}^{3}\textrm{r}\textrm{d}^{3}\textrm{r}^{^{\prime
}}\!\hat{\Psi}^{\dag }(\textrm{\textbf{r}})\hat{\Psi}^{\dag
}(\textrm{\textbf{r}}^{^{\prime
}}\!)V_{inter}(\textrm{\textbf{r}}\!\!-\!\!\textrm{\textbf{r}}^{^{\prime
}}\!)\!\hat{\Psi}(\textrm{\textbf{r}}^{^{\prime
}}\!)\hat{\Psi}(\textrm{\textbf{r}}), \label{manyh}
\end{eqnarray}%
where $\hat{\Psi}^{\dag }(\textrm{\textbf{r}})$,
$\hat{\Psi}(\textrm{\textbf{r}})$ are the creation and
annihilation field operators for the atomic field, respectively. $\hat{\mathbf{E}}%
(\textrm{\textbf{r}})$ is the electric field operator of the electromagnetic field, and $\hat{%
\mathbf{D}}(\textrm{\textbf{r}})$ is the atomic dipole moment
operator.
$V_{inter}(\textrm{\textbf{r}}-\textrm{\textbf{r}}^{^{\prime }})$
represents the interatomic potential. The counter-propagating light
forms a standing wave in the triangle ring cavity. The atoms in the
standing light feel a periodical potential. The condensate is
transversely tightly bounded, therefore its transverse freedom is
frozen, and it is reasonable to recognize the condensate as
quasi-one-dimensional ensemble in its longitudinal direction
\cite{Victor}. This suggests that the matter field operators can be
expanded with a set of Bloch functions, which are
amplitude-modulated plane waves,
\begin{eqnarray}
\hat{\Psi}(\mathbf{x})
&=&\sum\limits_{n}\hat{b}_{n}e^{i\mathbf{k}_{n}
\mathbf{x}}u(\mathbf{x}),  \notag \\
\hat{\Psi}^{\dag }(\mathbf{x}) &=&\sum\limits_{n}\hat{b}_{n}^{\dag
}e^{-i\mathbf{k} _{n}\mathbf{x}}u^{\ast }(\mathbf{x}), \label{field}
\end{eqnarray}%
where $u(\mathbf{x})$ is a periodical function along the direction
the horizontal axis of the cavity, $u(x+\frac{\lambda }{2})=u(x)$,
with the period one-half of the wave-length of the light wave.
$\hat{b}^{\dag}_{n}$ and $\hat{b_{n}}$ are the creation and
annihilation operators of the bosonic atoms in the $n$th Bloch mode.
$\textbf{k}_{n}=2n\textbf{k}_{R}$ is the wave vector of the
modulated plane wave. For the absorption and subsequently stimulated
emission of photons from the two modes of optical fields, the atoms
would acquire a net momentum of even number multiples of photon
momentum, $2n\hbar k_{R}$. Assuming the atoms are initially in the
stationary state,
after the scattering process, they will be excited to the higher adjacent momentum modes, which can be described as $(e^{2i%
\mathbf{k}_{R}\mathbf{x}}+e^{-2i\mathbf{k}_{R}\mathbf{x}})u_{0}(\mathbf{x})$,
with $u_{0}(\textbf{x})$ the envelop of the stationary wave packet.
If the atoms get further scattered, some of them may go to the even
higher momentum mode. For simplicity, we can approximate the
amplitude-modulated plane waves
with the wave functions of the free particles, and the wave functions of the atomic field now read,
\begin{eqnarray}
{|\Psi }(x)\rangle  &=&\sum\limits_{n}{\Psi }_{n}\left\vert \frac{1}{\sqrt{V}%
}exp(\frac{i}{\hbar }{p_{n}\cdot x})\right\rangle =\sum\limits_{n}{\Psi }%
_{n}|p_{n}\rangle,   \notag \\
{\langle \Psi }(x)| &=&\!\!\!\!\sum\limits_{n}{\Psi }_{n}^{\ast }\left\langle \frac{1%
}{\sqrt{V}}exp(\frac{-i}{\hbar }{p_{n}\cdot x})\right\vert \!\!=\!\!\sum\limits_{n}{%
\Psi }_{n}^{\ast }\langle p_{n}|.
\label{fieldp}
\end{eqnarray}%
In our case, due to the diluteness of the atomic gas, we are going
to neglect the modification of wave functions from the atomic
interation. Here, $V$ is the volume of the condensate, $\langle
x|p_{n}\rangle =\frac{1}{\sqrt{V}}\textrm{exp}(\frac{i}{\hbar
}{p_{n}\cdot x})$ is the wave function of free particles with
momentum $p_{n}=2n\hbar k_{R}$ in coordinate space. The particle
number of the atoms is normalized to one. At the same time, we have
removed the hat from each operator of the matter field, using a
mean-field treatment for the atomic field.

We approximate that the discrete modes of the atomic field makes a
complete set describing dynamics of the condensate. Plugging
Eq.~(\ref{fieldp}) into Eq.~(\ref{manyh}), we arrive at the
following expression for the Hamiltonian of the light-atom coupling
system,
\begin{eqnarray}
\hat{H}\!\!\!\!&=&\!\!\!\!\!\!\!\!\sum
\limits_{m=-\infty}^{\infty}\hbar\omega_{m}\left\vert
p_{m}\right\rangle \left\langle p_{m}\right\vert +\hbar G
\notag \\
&\cdot &\!\!\!\!\!\!\!\sum\limits_{n=-\infty}^{\infty}\!\!\!\!\!
\left(a_{k_{L}}^{\dag}a_{k_{R} }\left\vert p_{n}\right\rangle
\left\langle p_{n-1}\right\vert +a_{k_{R}
}^{\dag}a_{k_{L}}\left\vert p_{n}\right\rangle \left\langle p_{n+1}
\right\vert \right),  \label{hquan}
\end{eqnarray}
where we have adopted the full quantized version for the electric component of the light fields, $\mathbf{E}(\mathbf{r})=\sum\limits_{\mathbf{k}%
}\hat{\epsilon}_{\mathbf{k}}\mathscr{E}_{\mathbf{k}}(a_{\mathbf{k}
}+a^{\dag}_{\mathbf{k}}),\mathbf{k}=k_{L,R}$, in which $\hat{\epsilon}_{%
\mathbf{k}}$ is the unit polarization vector; $\mathscr{E}_{\mathbf{k}}=%
\sqrt{\hbar \omega_{c}/2\epsilon V_{c}}$ has the dimension of an
electric field with $V_{c}$ the quantization volume of the
macro-cavity \cite{Q-Optics}. $a_{\textbf{k}}$ and
$a^{\dag}_{\textbf{k}}$ are the creation and annihilation operators
of photons in the corresponding light mode, respectively. $\hbar
\omega _{m}=2(\hbar mk_{L})^{2}/M $ is the energy of the $m$th
unperturbed discretized mode, and $M$ is the mass of a single atom
within the condensate. $G=-\Omega^2/\triangle$ is the effective
two-photon atom-field coupling constant, with
$\Omega=-\frac{\mathbf{d}\cdot \hat{
\epsilon}_{\mathbf{k}}\mathscr{E}_{\mathbf{k}}}{\hbar}$ the
single-photon atom-field coupling constant and $\mathbf{d}$ the
electric dipole moment of the atom. The form of $G$ comes from the
adiabatic elimination of the internal excited state of the two-level
atom, and this is legitimated by large detuning of the light fields.
This Hamiltonian is rather reminiscent of an ordinary lattice model
in solid state physics in spite of an atomic interacting term. The
second hopping term disturbs the motion of the free atoms and
induces exchanging of particles between different diffraction
orders. Such atomic diffractions are categorized into Raman-Nath
diffractions with short pulse fields and Bragg diffractions with
longer time of light-atom interaction.

To investigate the atomic motion of the system, it is usually convenient to
work in the interaction picture by seperating the Hamiltonian (\ref{hquan})
into the free part and the atom-field interacting part, then one has,
\begin{eqnarray}
\hat{\mathscr{V}}\left( t\right) & = &\hbar G\sum\limits_{n=-\infty}^{\infty
}\{ a_{k_{L}}^{\dag} a_{k_{R}} \left\vert p_{n}\right\rangle \left\langle
p_{n-1}\right\vert e^{-i\delta_{-}\left( t\right) }  \notag \\
&+&a_{k_{R}}^{\dag}a_{k_{L}}\left\vert p_{n}\right\rangle \left\langle
p_{n+1}\right\vert e^{-i\delta_{+}\left( t\right) }\},  \label{interh}
\end{eqnarray}
where $\delta _{\pm }=\pm16\pi\nu_{r}(n\pm\frac{1}{2})$, and
$\nu_{r}=\hbar k_{L}^{2}/(4\pi M)$ is the atomic one-photon recoil
frequency ($\nu_{r} \sim 3.77$ kHz according to the parameters
chosen in this system).

The optical freedom of this system can be integrated out to extract
the effective information of atomic dynamics. Unlike the classical treatment of the optical fields in which
the optical fields are recognized as plane waves with slowly varying amplitudes, the atomic motion of the system can
be explicitly modified by the status of the quantized light fields,
not just by the strength of the pump beams \cite{Li Ke,zhang}.
To see this, we shall assume, for example, that the light fields are
in number states. The Hamiltonian (\ref{interh}) conserves the total
photon number of the system. Thus the wave function of the
matter-wave condensate can be written as a linear combination of
different momentum states with the corresponding photon
distributions among the two light fields,
\begin{equation*}
\left\vert \Psi \left( t\right) \right\rangle =\sum\limits_{n}\Psi
_{n}\left\vert \psi \left( n,N_{R}-n,N_{L}+n,t\right) \right\rangle
,
\end{equation*}%
where $n=0,\pm 1,\pm 2,$ refers to the diffraction order. $N_{R}$
and $N_{L}$ are the initial photon number distribution among the two
light modes. $\left\vert \psi \left( n,N_{R}-n,N_{L}+n,t\right)
\right\rangle $ is the product of the wave function of the atomic
center-of-mass motion and that of the optical fields, $|p_{n}\rangle
\otimes |\psi _{f}\rangle $. In this stage, $|\psi _{f}\rangle $ is
choosen to be the number state noted as $|n_{1},n_{2}\rangle $
corresponding to that there are $n_{1}$ photons in the right-going
light mode and $n_{2}$ photons in the left-going mode. Since the
condensate is initially prepared in the stationary state, each atom
needs to absorb $n$ photons from the right-going light mode in order
to hop to the $n$th momentum mode. Therefore, the photon number
distribution of the two light modes is directly related to the
atomic diffraction order. The equation of atomic motion now reads,
\begin{eqnarray*}
i \hbar &\dfrac{\partial }{\partial t}&\left\vert \Psi \left( t\right)
\right\rangle =\hat{\mathscr{V}}\left( t\right) \left\vert \Psi \left(
t\right) \right\rangle  \nonumber \\
&=&\hbar G\sum\limits_{n=-\infty }^{\infty }\{a_{k_{L}}^{\dag
}a_{k_{R}}\left\vert p_{n}\right\rangle \left\langle p_{n-1}\right\vert
e^{-i\delta _{-}\left( t\right) }  \nonumber \\
&&+a_{k_{R}}^{\dag }a_{k_{L}}\left\vert p_{n}\right\rangle \left\langle
p_{n+1}\right\vert e^{-i\delta _{+}\left( t\right) }\}\left\vert \Psi \left(
t\right) \right\rangle .
\label{motion}
\end{eqnarray*}
Keeping the photon degree of freedom, the above equation can be further simplified to,
\begin{eqnarray*}
&\dot{\Psi}_{n}(t)&\left\vert n_{R}-n,n_{L}+n\right\rangle
\nonumber\\
&=&a_{k_{L}}^{\dag
}a_{k_{R}}\Psi _{n-1}e^{-i\delta _{-}t}\left\vert
n_{R}-n+1,n_{L}+n-1\right\rangle \\
&&+a_{k_{R}}^{\dag }a_{k_{L}}\Psi _{n+1}e^{-i\delta _{+}t}\left\vert
n_{R}-n-1,n_{L}+n+1\right\rangle ,
\end{eqnarray*}
Until now we have kept quantum character of the optical fields.
Integrating the photon degrees of freedom by explicitly executing
the operations of the photon creation and annihilation operators on
the corresponding number states of the light fields, we arrive at
the following equation of the atomic motion,
\begin{eqnarray}
\dot{\Psi}_{n}(t)=W_{R}^{n}\cdot \Psi _{n-1}e^{-i\delta _{-}t}+W_{L}^{n}\cdot \Psi
_{n+1}e^{-i\delta _{+}t},
\label{motionq}
\end{eqnarray}
where $W^{n}_{L,R}$ are the forward and backward transition weights
defined as, $ W_{R}^{n}=-iG\sqrt{\left( n_{R}-n+1\right) (n_{L}+n)}$
and $W_{L}^{n}=-iG\sqrt{ (n_{R}-n)\left( n_{L}+n+1\right) }$. This
equation has included the quantum effect of optical fields upon the
atomic motion. The complicated expressions for the transition
weights are resulted from the property of number state of the
optical fields that when it is operated by the corresponding
creation and annihilation operators, we have
$a^{\dag}|n\rangle=\sqrt{n+1}|n+1\rangle$ and
$a|n\rangle=\sqrt{n}|n-1\rangle$. It can be directly observed from
Eq.~(\ref{motionq}) that, the transition weights depend not only on
the diffraction order but also on the photon distribution among the
two light modes. Thus, unlike the classical case of plane wave
approximation for the light fields, there is no definite equality
between $W^{n}_{R}$ and $W^{n}_{L}$; $|W^{n}_{R}|>|W^{n}_{L}|$ and
$|W^{n}_{R}|<|W^{n}_{L}|$ are both possible, depending on whether
$N_{R}<N_{L}$ or $N_{L}>N_{R}$. Table (\ref{tab1}) lists out the
transition weights of the first few order atomic scattering
processes for $N_{R}=80$ and $N_{L}=10$. It shows that for the first
few order scattering processes, $|W^{n}_{R}|<|W^{n}_{L}|$ are always
satisfied.

\begin{table}[!h]
\begin{tabular}{|c|c|c|c|}
\hline
 Scattering Process & $|W^{n}_{R}|$ & $|W^{n}_{L}|$ & $ |W^{n}_{R}\cdot W^{n}_{L}|$\\ \hline
$|p_{0}\rangle\rightarrow|p_{1}\rangle,|p_{-1}\rangle$ & $28.46$ & $29.66$ & $ 844.12$ \\ \hline
$|p_{1}\rangle\rightarrow|p_{0}\rangle,|p_{2}\rangle$ & $29.66$ & $30.78$ & $ 912.93$ \\ \hline
$|p_{-1}\rangle\rightarrow|p_{0}\rangle,|p_{-2}\rangle$ & $27.16$ & $28.46$ & $ 772.97$ \\ \hline
\end{tabular}
\caption{The transition weights in dependence on the diffraction
order in unit of atom-field coupling constant $G$, for the initial
photon number distribution, $N_{R}=80$, $N_{L}=10 $.}
\label{tab1}\tabcolsep 10pt \centering
\end{table}

\begin{figure}[!t]
\begin{center}
\scalebox{0.50}[0.52]{\includegraphics*[60pt,90pt][450pt,530pt]{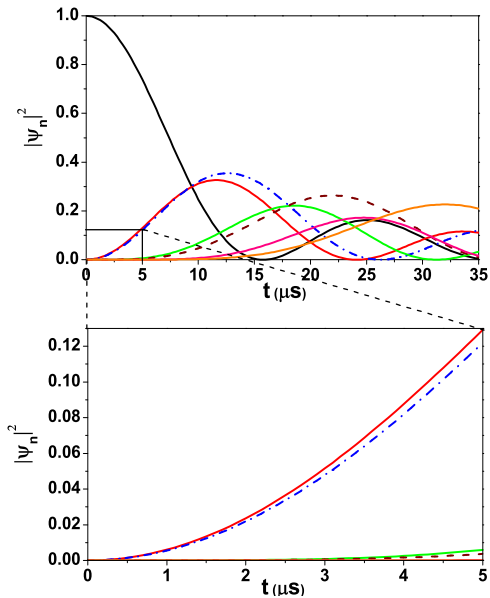}}
\end{center}
\caption{(Color online). Atomic probability distribution of the
first few discrete momentum modes in the limit of short light-atom
interacting time, for an initial photon number distribution,
$N_{R}=80, N_{L}=10$.  Black solid curve refers to $n=0$ mode; red
solid curve , $n=1$; blue dash-dotted curve, $n=-1$; green-solid
curve, $n=2$; wine dashed curve $n=-2$; pink solid curve, $n=3$;
orange solid curve, $n=-3$. The light-atom coupling constant is
chosen $G=0.7\nu_{r}$, with $\nu_{r}$ the atomic one-photon recoil
frequency.} \label{Raman}
\end{figure}

\section{light fields in number states}
\subsection{Raman-Nath regime}

To solve Eq.~(\ref{motionq}) in the full time-domain, one has to
appeal to numerical calculations. Fortunately, in the Raman-Nath
regime where the atoms interact with the light fields for rather
short time, the low order diffractions dominant the whole atomic
scattering process, and Eq.~(\ref{motionq}) can be solved
analytically. In this limit $t\rightarrow 0$, the probability
equation of the atomic motion takes the following form,
\begin{eqnarray}
\dot{\Psi}_{n}(t)=W_{R}^{n}\cdot \Psi _{n-1}+W_{L}^{n}\cdot \Psi
_{n+1}.
\label{motionqR}
\end{eqnarray}
For macroscopic occupations of photons in the two light modes, the
diffraction order $n$ is small compared to the initial photon number
$N_{R}$ and $N_{L}$, i.e., $N_{R,L}\gg n$, we have $W_{R}^{n}\sim
W_{L}^{n}\sim \sqrt{N_{R}N_{L}}$ which can be observed in Table
(\ref{tab1}). Thus Eq.~(\ref{motionqR}) has the following solution,
\begin{equation}
\Psi _{n}\left(t\right) =i^{n}e^{i\theta}(\frac{W_{R}^{n}}{W_{L}^{n}})^{\frac{n%
}{2}}J_{n}\left( \xi t\right)
\label{asolution}
\end{equation}
where $\xi =2\sqrt{(W_{R}^{n}W_{L}^{n})}$, $J_{n}\left( \xi t\right)
$ is the $n$th order Bessel function; $\theta$ is an arbitrary phase
angle. The population of atoms occupying the $n$th momentum state
$\left\vert p_{n}\right\rangle $ is $\rho _{n}(t)=|{\Psi
_{n}(t)}|^{2}=(\frac{W_{R}^{n}}{W_{L}^{n}})^{n}[J_{n}(\xi t)]^{2}$.
For the inequality of $W_{R}^{n}$ and $W_{L}^{n}$ as show in
Table~(\ref{tab1}), the atomic diffraction will display exotic
behaviors. Fig.~\ref{Raman} shows the probability amplitudes of the
atoms in different momentum modes evolving with the light-atom
interacting time. It can be seen that, besides that the atomic
populations in different momentum modes oscillate with time, the
curves of the momentum oscillations of the $\pm n$th modes do not
overlap. Unlike the situations illustrated in \cite{Li Ke, zhang},
the probability amplitudes of these $\pm n$th momentum modes do not
reach the minimum or maximum values exactly at the same time. For
example in Fig.~\ref{Raman}, there is a shift between the first
minima of the probability amplitudes for the $\pm 1$st order atomic
momentum modes. Even though to study the atomic motion for
long-period light-atom interaction from the data derived in this
regime may not seem adequate enough, it implies that the atomic
diffractions would be asymmetric about the zeroth order, and such
asymmetries are induced by the number-state-nature of the optical
fields. More detailed discussion and the accuracy of the analytical
solution in this short time regime will be demonstrated in the later
part of this paper.

\subsection{Bragg regime}

In the Bragg diffraction regime, where the light and atoms
interact with each other for a longer time, the exponential factors
in Eq.~(\ref{motionq}) greatly modify the atomic motion, compared to
the situations of the short time light-atom interactions (i.e.,
Raman-Nath regime). However, in this regime, the equation of atomic
motion cannot be explicitly solved analytically. To obtain the
information of the atomic motion in the full-time domain, numerical
calculation is carried out with a cutoff at the the $n=\pm10$th
momentum modes in order to maintain the accuracy. In fact, the
validity of such a cutoff has been implied in the case of short-time
light-atom interactions; for the very high diffraction orders, the
atomic occupations in the relevant modes are always vanishingly
small.

\begin{figure}[!t]
\begin{center}
\scalebox{0.80}[0.82]{\includegraphics*[45pt,0pt][415pt,575pt]{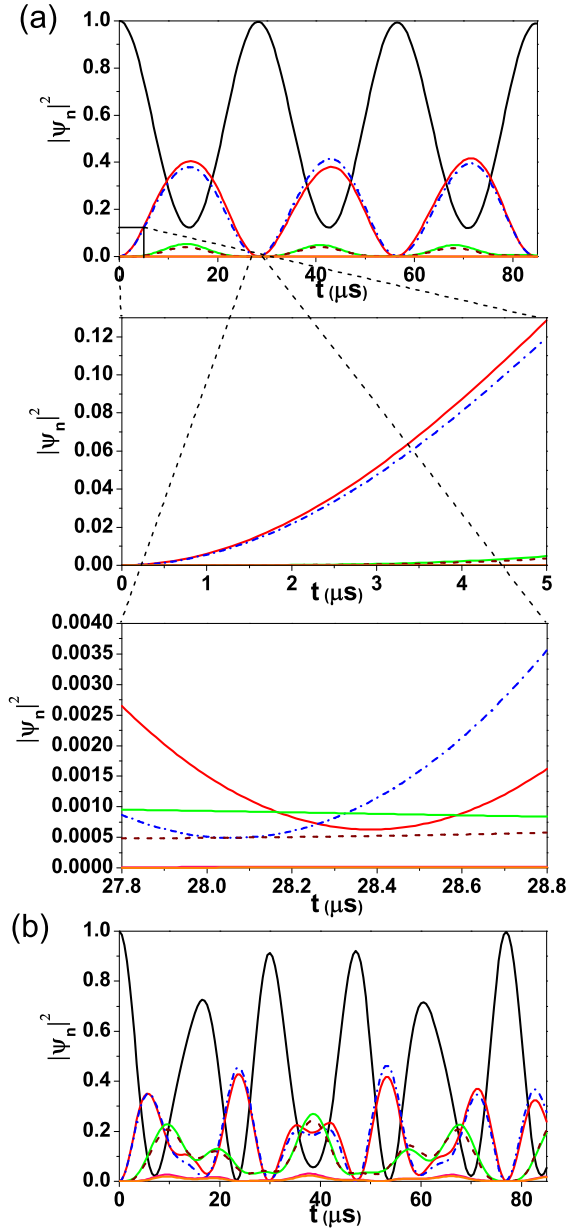}}
\end{center}
\caption{(Color online). Atomic probability distribution on the
first few discrete momentum modes in the regime of long light-atom
interacting time, for the initial photon number distribution, (a)
$N_{R}=80, N_{L}=10$; (b) $N_{R}=80, N_{L}=10$. [Line styles are the
same with that in Fig \protect\ref{Raman}(a)]} \label{quantum}
\end{figure}

Because of the dependence of the transition weights upon the
diffraction order and the photon number distribution, numerical
simulation displays interesting behaviors of the atomic motion in
the quantized light fields. It seems that the atoms can see the
imbalanced intensities of optical fields, and thus take actions
accordingly. Fig.~\ref{quantum} shows the probability amplitudes of
atoms in the first few order momentum modes for different photon
number distributions. It is found that, the quasi-period of momentum
oscillations is changing with respect to the diffraction order.
Fig.~\ref{quantum}(a) shows the probability amplitudes of the
scattered atoms at the photon number distribution
$N_{R}=80,N_{L}=10$. Even for atomic diffractions of the same order,
see the $n=\pm1$ curves, the atomic populations oscillate at
different frequencies. At $t\approx 28\mu s$, the curves for the
$n=\pm 1$ modes get to their lowest points, but if one goes to the
details in this interval, it is found that there is a shift between
the minima of two curves. That is to say the populations of atoms in
these two modes are in fact oscillating with different periods.
Additionally, it is found that the atoms would favor the forward and
backward diffractions alternatively; the $n=1$ order is sometimes
stronger and sometimes weaker than the $n=-1$ order, implying that
asymmetry appears in the matter-wave diffractions. This peculiar
phenomenon is resulted from the quantum nature of the optical
fields. The asymmetric atomic diffractions can be manipulated by
adjusting the imbalance of the photon number distributions in the
two light modes. Fig. \ref{quantum}(b) shows the atomic distribution
of the diffracted atoms for $N_{R}=200,N_{L}=20$. Similar
diffraction behavior is observed as that shown in Fig.
\ref{quantum}(a). However, the atomic distributions in this case are
oscillating at higher frequencies. Thus the imbalanced photon number
distribution of the light fields modifies the dynamics of the atomic
diffraction, and the order-dependence of the transition weights
induced by the number-state-nature of the light fields makes the
atomic momentum oscillation very irregular.

The above result is quite different from that obtained from the
traditional analysis, where mean-field approximations are applied to
both the atomic field and optical fields. The photon creation and
annihilation operators are directly replaced with their mean-field
values. Even though this is efficient for extracting qualitative
information about the atomic dynamics, it ignores the details of
influences from the quantum nature of the light fields upon the
atomic motion.

\begin{figure}[!t]
\begin{center}
\scalebox{0.56}[0.60]{\includegraphics*[20pt,60pt][380pt,490pt]{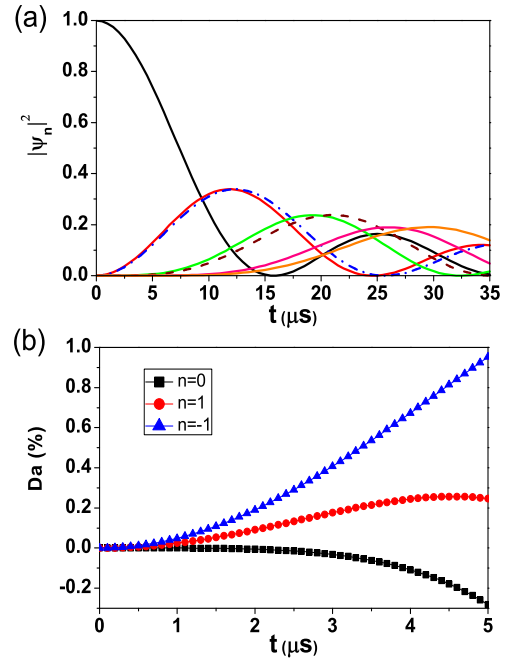}}
\end{center}
\caption{(Color online). (a) Atomic probability distribution of the
first few discrete momentum modes from numerical calculation in the
Raman-Nath regime, for the photon number distribution
$N_{R}=80,N_{L}=10$. [Line styles are the same with that in Fig
\ref{Raman}(a)]. (b) The error of the analytical solution obtained
in the short time limit of the light-atom interaction, compared to
that obtained in the Bragg regime.} \label{accu}
\end{figure}

To check the accuracy of solution derived in the short time limit of
the light-atom interactions in the last section, numerical
simulation for Eq.~(\ref{motionqR}) is carried out for the photon
number distribution $N_{R}=80,N_{L}=10$, and the result is shown in
Fig.~\ref{accu}(a). It is found that the data obtained here is
consistent with that obtained in the Bragg regime. There are
displacements between the minima (or maxima) of the conjugate $\pm
n$ curves, and the probability amplitude of the $n=1$ momentum mode
can be either larger or smaller than that of the $n=-1$ mode. This
is similar to the result displayed in Fig.~\ref{quantum}(a).
However, we have to emphasize that, using Eq.~(\ref{motionqR}) to
describe atomic motion is only valid in the Raman-Nath regime with
short time light-atom interactions; thus data in Fig.~\ref{accu}(a)
for large $t$ is not trustful. Fig.~\ref{accu}(b) shows the error of
the data in this regime compared to that from Fig.~\ref{quantum}(a).
The accuracy function $D_{a}$ is defined as
$D_{a}^{n}(t)=\frac{|\Psi_{n}^{R}|^{2}-|\Psi_{n}^{B}|^{2}}{|\Psi_{n}^{R}|^{2}+|\Psi_{n}^{B}|^{2}}$,
where $\Psi_{n}^{R}$ is the probability amplitude of the atomic
motion in the Raman-Nath regime obtained in section II.~A, and
$\Psi_{n}^{B}$ is that obtained in the Bragg regime. It is found
that, for short time, $t<5\mu s$, the error of the analytical
solution Eq.~(\ref{asolution}) is less than $1\%$. So it is a good
approximation of the atomic motion in this regime.

\section{light fields in coherent states}
\begin{figure}[!t]
\begin{center}
\scalebox{0.51}[0.55]{\includegraphics*[80pt,305pt][440pt,530pt]{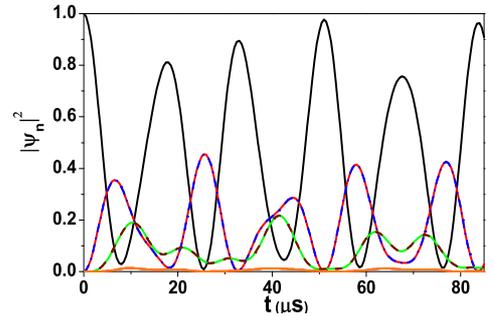}}
\end{center}
\caption{(Color online). Atomic probability distribution of the
first few discrete momentum modes in the condition that the light
fields are in coherent sates, $|\alpha_{R}\rangle$ and
$|\alpha_{L}\rangle$. The parameters are chosen such that
$W_{co}=\alpha_{R}\alpha_{L}^{*}=60+40i$. [Line styles are the same
with that in Fig \ref{Raman}(a)].} \label{coherent}
\end{figure}

Besides that the properties of atomic diffractions are sensitively
dependent on the intensities of the pump fields, the status of the
light fields can also impose implicit corrections to the atomic
diffractions. To make comparison to the case that the lights are in
number states, we turn to the case that the two optical fields are
in coherent sates $|\alpha_{R}\rangle$ and $|\alpha_{L}\rangle$,
respectively. $\alpha_{R}$ and $\alpha_{L}$ are the eigenvalues of
the annihilation operators of the two light modes on the relevant
coherent states. The equation of the atomic motion reduces to
\begin{eqnarray}
\dot{\Psi}_{n}(t)=W_{co}\cdot \Psi _{n-1}e^{-i\delta _{-}t}+W_{co}^{*}\cdot \Psi
_{n+1}e^{-i\delta _{+}t},
\label{motionqC}
\end{eqnarray}
where $W_{co}$ and $W_{co}^{*}$ are conjugate with each other,
$W_{co}=\alpha_{R}\alpha_{L}^{*}$. Generally, $\alpha_{R}$ and
$\alpha_{L}$ are complex numbers. If $\alpha_{R}$ and $\alpha_{L}$
are chosen to be real, then $W_{co}=W_{co}^{*}=\alpha_{R}\alpha_{L}$
and the equation is of the same form as that obtained in \cite{Li
Ke,zhang}. By numerical simulation, the information of the atomic
motion is obtained in the full-time domain. Fig.~\ref{coherent}
shows the probability amplitudes of the first few momentum modes at
$W_{co}=60+40i$. It is found that the curves of the $\pm n$th orders
coincide with each other; thus the forward and backward atomic
transitions are equally favored, and no asymmetry appears in the
atomic diffraction. This can be seen from the fact that the absolute
values of transition weights for the forward and backward
transitions are the same, and their phase difference can be absorbed
in the definition of the phase factor $\delta_{\pm}$ as constant,
thus the equation of the atomic motion still takes the symmetric
form as that in \cite{Li Ke,zhang}.

In this case, the transition weights are independent upon the
diffraction order. The data obtained here resembles that with
classical treatments of the optical fields, where the
electromagnetic field are recognized as plane waves. Comparing the
results with that in the case where the optical fields are in number
states, even though the equation of the atomic motion is also
derived in a full quantum treatment of the light fields, only
symmetric atomic diffraction can be obtained here; this is
determined by the specific status of quantized light fields.

\section{Summary}

In this paper, we investigate the atomic diffractions in quantized
light fields. The quantum nature of the light fields induces exotic
phenomena in the atomic scattering processes. For light fields in
number states, the atomic transitions among different discrete
momentum modes depend on both the transition order and the initial
photon number distribution. The quasi-periods of the atomic momentum
oscillations are implicitly modified, and the atomic diffraction is
no longer symmetric. However, for light fields in coherent states,
the atomic diffractions are still symmetric, this verifies the
coherent-state-approximation of the light fields which is frequently
utilized in the cavity quantum electrodynamics. Situations of other
kind of states of quantized light fields, such as arbitrary linear
combination of number states, can also be investigated using the
methods introduced in this paper. The results obtained in this paper
can be extended to other light-atom interacting systems for enabling
new physics in atom optics.

We would thank Dr. L. Deng for stimulating discussions and
encouragement. This work is supported by NSFC under grants Nos.
10874235, 10934010, 60978019, the NKBRSFC under grants Nos.
2009CB930701, 2010CB922904, and 2011CB921502, and NSFC-RGC under
grants Nos. 11061160490 and 1386-N-HKU748/10.

\end{document}